\documentstyle[aps,epsfig]{revtex}

\makeatletter

\begin{document}

{\par\centering \textbf{\Large Angular Correlation in Double Photoionization
of Atoms and the Role of the Detection Process }\Large \par}

\vspace{0.15in}
{\par\centering \textbf{\large Dipankar Chattarji and Chiranjib Sur }\large \par}
\vspace{0.1in}

{\par\centering Department of Physics, Visva-Bharati, Santiniketan 731 235, INDIA\emph{ }\par}
\vspace{0.2in}

{\small The problem of angular correlation in the double photoionization (DPI)
of rare gas atoms is considered in some depth. We refer particularly to the
efficiency operator for the detection of an electron by a detector having cylindrical
symmetry. The different factors in the efficiency operator are discussed in
detail keeping in mind the fundamental epistemological question of the role
of the detection process in such experiments.}{\small \par}

\textbf{\small PACS No} {\small : 32.80.H, 32.80.F, 03.65.T,79.20.F }{\small \par}
\vspace{0.2in}

In this paper we wish to consider the problem of angular correlation between
the two electrons emitted by an atom when it is doubly ionized by a photon.

Consider a randomly oriented rare gas atom in a \( ^{1}S^{e} \) state. The
atom absorbs an unpolarized photon and after a certain time interval emits a
photo-electron from one of the inner shells giving a singly ionized atomic state.
This intermediate ionic state now de-excites by emitting an Auger electron,
typically from an outer shell, giving rise to a two-vacancy final atomic state.
We can denote this sequence of events as follows.

\begin{equation}
\label{one}
h\nu +\mathbf{A}\longrightarrow \mathbf{A}^{+}+e_{1}^{-}\longrightarrow \mathbf{A}^{++}+e_{1}^{-}+e_{2}^{-}.
\end{equation}
 A polar plot of the observed distribution of coincidences between the two emitted
electrons as a function of the angle~ between their directions of emission shows
a clear periodic behaviour {[}\ref{schmidt}{]}. The question we ask is: what
is the origin of this angular correlation? Could it have anything to do at all
with the process of detecting the electrons? On the face of it, this last
question may not seem so obvious. It will, however, become clearer as we proceed
with our discussion.

Double photoionization (DPI) occurs when an atomic target like the one described
above is irradiated with a monochromatic photon beam from an advanced light
source, e.g. a synchrotron. Along with single photoionization (PI), there may
be events in which~ two electrons are emitted by an atom in quick succession.
In case the time interval between the successive emission of the two electrons
is substantially longer than the time taken by the first electron to leave the
interaction zone, DPI may be regarded as a two-step process {[}\ref{we1}{]}.
In other words, the emission of the two electrons may be regarded as being clearly
separated in time. This in its turn will depend on the energy imparted to the
atom by the incident photon.

We wish to obtain an angular correlation function for the two emitted electrons
in terms of the angle between their directions of emission. We shall do this
by considering an ensemble of such atomic systems belonging to all possible
quantum mechanical states \( Q \). Each state \( Q \) is labeled by the total
angular momentum \( J \), its projection \( M \), and the remaining set of
quantum numbers \( \alpha  \).

The angular correlation function \( W(\theta ) \) for the two emitted electrons
is the probability that the angle between their directions of emission is \( \theta  \).
Evidently this is a statistical quantity, and \( W(\theta ) \) would have to
be the ensemble average of the above probability.

Now, how do we determine this probability? Hopefully, we let the atomic system
attain the final state given in Eq.(\ref{one}), we set up two detectors at
a suitable distance from the reaction zone with their axes making an angle \( \theta ^{\prime } \)
with each other, and we try to detect coincidences between the photo- and Auger
electrons. The number of coincidences we can hope to detect will depend on two
distinct factors.

(i) There is a certain probability for the atomic system to attain the final
state. This is described by the appropriate matrix element of the density or
statistical operator \( \rho  \) {[}\ref{fano},\ref{blum}{]}.

(ii) Even if the system goes over to the final state, because of the finite
size of the detectors and other important limiting factors, a coincidence event
may or may not be detected. There is thus a finite probability \( \epsilon \, (0\leq \epsilon \leq 1) \)
that the event will be detected by the system of detectors. This probability is represented by the efficiency
operator \( \varepsilon  \). It will depend on the size, position and geometrical
configuration of the detectors, but not their internal physical or chemical
nature provided that there is full absorption of an electron within the material
of a detector {[}\ref{rose}{]}. 

Obviously, \( W(\theta ) \) will be given by the joint probability of the formation
of the final state and its detection by the system of detectors, i.e. by the
product of \( \rho  \) and \( \varepsilon  \). For a given state \( Q \)
this joint probability will be \( \rho \varepsilon  \) . The average probability
\( \overline{\varepsilon } \) for the ensemble will be given by the trace of
the product matrix. We write

\begin{equation}
\label{two}
\begin{array}{cc}
\overline{\varepsilon } & =\sum _{Q}\varepsilon _{Q}\left\langle Q\right| \rho \left| Q\right\rangle \\
 & =\sum _{Q}\varepsilon _{Q}\rho _{QQ}\\
 & =Tr(\varepsilon \rho )\\
 & =Tr(\rho \varepsilon )\, .
\end{array}
\end{equation}
 Here \( \varepsilon _{Q} \) is the efficiency or probability of detection
of the state described by the quantum numbers \( Q \), and \( \rho _{QQ} \)
the probability of the system being in the particular state \( Q \) {[}\ref{coester}{]}.
Both \( \varepsilon  \) and \( \rho  \) are tensor operators.

Since the angular correlation function happens to be the trace of a matrix {[}\ref{coester},\ref{ferguson}{]},
it will be invariant under a unitary transformation in Hilbert space.

Another property of the system arises from the random orientation of the rare
gas atoms. The electrons emitted by them are unpolarized. And we take the detectors
to be insensitive to electron polarization. Hence the angular correlation function
itself can depend only on scalar invariants formed of the unit momentum vectors
of the two emitted electrons \( \widehat{\mathbf{p}_{1}} \) and \( \widehat{\mathbf{p}_{2}} \).
These invariants are given by the scalar product of spherical tensors {[}\ref{satchler}{]},
as follows:

\begin{equation}
\label{three}
\begin{array}{cc}
\mathbf{C}_{k}(\widehat{\mathbf{p}_{1}})\cdot \mathbf{C}_{k}(\widehat{\mathbf{p}_{2}}) & =\sum _{m}C_{km}(\widehat{\mathbf{p}_{1}})C^{\star }_{km}(\widehat{\mathbf{p}_{2}})\\
 & =P_{k}(\widehat{\mathbf{p}_{1}}\cdot \widehat{\mathbf{p}_{2}})=P_{k}(cos\theta )\, .
\end{array}
\end{equation}
 In Eq.(\ref{three}) \( P_{k}(cos\theta ) \) is a Legendre polynomial. The
index \( k \) will be restricted to the allowed values of the resultant of
the angular momenta \( j_{1} \) and \( j_{2} \) of the two emitted electrons.
Out of these, odd values of \( k \) will drop out because they would give odd
parity.

Now, from the elements of statistical mechanics, we know that \( \overline{\varepsilon } \)
is the expectation value (or average value) of the efficiency operator \( \varepsilon  \)
{[}\ref{terharr}{]}. It needs to be pointed out that this expectation value
is a function of \( \theta  \). Thus the angular correlation function \( W(\theta ) \)
is, to within a multiplying factor, just the expectation value of the efficiency
operator for a given value of \( \theta  \). To be more precise, it represents
the angle-dependent factor in the expectation value~ of the efficiency operator
for the entire detecting system.

Now, based on physical considerations, can we find an expression for the efficiency
operator?

We begin by noting that the efficiency operator for a single electron represents
the attenuation of the probability of detecting the signal 
caused by certain geometrical properties of the detector as well as certain
intrinsic limitations of the detection process to be discussed later. Let us
now try to write down an expression for the efficiency operator for a single
detector in detecting an electron in terms of such factors.

Obviously, a co-ordinate representation would be the most appropriate for the
discussion of these factors. But what kind of co-ordinate system shall we use?

Because of the spherical symmetry of the system in the interaction region, we
use spherical polar co-ordinates for our calculation {[}\ref{we2}{]}. However,
as soon as an emitted electron begins to interact with a detector, it will acquire
a symmetry appropriate to the detector. In the present paper our objective is
to examine the geometrical properties of the electron detector and to see how
they affect its efficiency. So let us see what kind of symmetry exists in the
detection region.

Let us start by considering our options with regard to the shape and size of
the detector. Could we, for example, start with the limiting case of a point
detector {[}\ref{ferguson32}{]}? By considering the signal to noise ratio,
it may be shown that one must use a finite size detector. In order to find the
direction of emission of an electron it must have axial symmetry. Furthermore,
the photo-electron as well as the Auger electron is characterised by a well
defined energy and a well defined orbital angular momentum. The well defined
energy implies that we need an electron spectrometer which is a differential
energy analyser. And the well defined orbital angular momentum indicates that
the spectrometer should be an angle integrated device. The cylindrical mirror
analyser (CMA) meets all these requirements {[}\ref{URL},\ref{dc}{]}. Hence
it seems to be the obvious candidate. Recent angular correlation measurements
{[}\ref{schmidt2}{]} using the CMA clearly support this view.

A CMA receives incident electrons through a circular aperture of finite radius
\( r \). Hence the angle of incidence \( \beta _{i} \) of an electron as measured
with respect to the cylinder axis varies from \( 0 \) to a small finite value.
Here the index \( i=1 \) for photo-electrons and \( 2 \) for Auger electrons.
 But, whatever its
actual value, because of the angle integrated character of the CMA the observer
has no way of knowing \( \beta _{i} \). He identifies the direction of emission
of the electron with the axis of the CMA.

As a result of the identification of the direction of emission of the electron
with the cylinder axis there is an effective rotation \( \Re _{i}(=0\beta _{i}0) \)
of its direction of emission. Thus the efficiency operator representing the detection process
will contain a rotation matrix \( D_{\kappa ^{\prime }_{i}\kappa _{i}}^{k_{i}}(\Re _{i}) \).
This rotation matrix will be a factor in the expression for the efficiency operator of a detector.

In our paper under reference {[}\ref{we2}{]} we have shown that the expression
for the efficiency operator of a single detector contains a rotation matrix
element. It will be seen from the formal derivation given there that the angle-dependence
of the angular correlation function arises directly from this rotation
matrix. 

Note that the angle integrated character of the electron spectrometer is a requirement
imposed by the fact that it has to receive electrons with a well defined orbital
angular momentum. In other words, the complete indeterminacy in the direction
of emission of an electron is a dynamical requirement and not a matter of technical
deficiency of the detector. Since the rotation matrix in the expression for
the angular correlation function originates from this indeterminacy, evidently
the angular correlation arises directly from it. 

This takes care of the most important factor in the expression for the efficiency
operator, namely the rotation matrix element. We shall call this the attenuation
factor due to rotation. Before we go on to the other factors, we take a brief
look at the geometrical arrangement of the detectors.

In Fig.1, \( A_{1} \) represents the axis of the detector set up to detect the photo-electron,
and \( A_{2} \) the axis of the detector receiving the Auger electron. 
The angle between the two axes is \( \theta ^{\prime } \). The directions
\( D_{1} \) and \( D_{2} \) are the actual directions of emission of the photo-
and Auger electrons respectively. The angle \(\beta_{1} \) is the angle between \(A_{1}\) and \(D_{1}\), i.e. the angle of incidence of the photo-electron as defined above. Similarly, \(\beta_{2}\), the angle between \(A_{2}\) and \(D_{2}\), is the angle of incidence of the Auger electron.

Each CMA has a circular aperture of radius \(r\) for receiving electrons. In a DPI experiment the base of each of the two CMAs used to detect the photo- and Auger electrons is placed at a distance \(h\) from the centre of the reaction zone.
The angular width of the aperture in each CMA as seen from the centre of
the target is \( 2\gamma  \), where \( tan\, \gamma =\frac{r}{h}\, (\sim .01) \).

We can now go back to the form of the efficiency operator corresponding to a
single detector detecting an electron. It is a tensor operator of rank \( k \)
with \( (2k+1) \) components. Following reference {[}\ref{we2}{]} we write
a reduced matrix element of the component labeled by \( \kappa  \) of the tensor
as

\begin{equation}
\label{four}
\varepsilon _{k\kappa }(jj^{\prime })=\sum _{\kappa ^{\prime }}z_{k}c_{k\kappa ^{\prime }}(jj^{\prime })D_{\kappa \kappa ^{\prime }}^{k}(\Re )\, .
\end{equation}
 Here \( \kappa  \) and \( \kappa ^{\prime } \) are projection quantum numbers
corresponding to the angular momentum \( k \). \( z_{k} \) is the attenuation
factor due to the finite size of a detector. It is different for different values
of \( k \). The factor \( c_{k\kappa ^{\prime }}(jj^{\prime }) \) arises from
the change of symmetry as the electron goes to the detection zone from the reaction
zone. Let us now look at these factors. 

(a) \emph{Attenuation due to absorption in a detector of finite size}. Let us
first consider the case of a single detector detecting, say, a photo-electron.
Here we can think in terms of an angular distribution measurement. The angular
distribution~ too can be written out as a Legendre polynomial expansion. The
attenuation factor multiplying \( P_{n}(cos\beta ) \) will be {[}\ref{rose}{]},

\begin{equation}
\label{five}
z_{n}=\frac{J_{n}}{J_{0}},
\end{equation}
 where

\begin{equation}
\label{six}
J_{n}=\int ^{\gamma }_{o}P_{n}(cos\beta )sin\beta d\beta \, .
\end{equation}

For an angular correlation experiment with two detectors having finite size
attenuation factors \( z_{n}(1) \) and \( z_{n}(2) \) the total attenuation
factor for the \( n \)th term will be

\begin{equation}
\label{seven}
Z_{n}=z_{n}(1)z_{n}(2)\, .
\end{equation}

(b) \emph{Attenuation factor corresponding to the state of polarization}. We
have already discussed the axial symmetry acquired by unpolarized electrons
as they enter the detection zone. What does this do to their quantum mechanical
state?

Let us first consider the semi-classical vector model. Axial symmetry about
the detector axis implies that the angular momentum vector of an electron can
only lie in the plane perpendicular to that axis, i.e. the \( xy \) plane.
Obviously, the \( z \)-component of its angular momentum will be zero. Now
going over to quantum mechanics, only those states will survive for which the projection quantum number \( \nu =0 \).
This calls for a projection operator having the form

\begin{equation}
\label{ten}
c_{n\nu }(jj^{\prime })=N_{jj^{\prime }n}C^{jj^{\prime }n}_{\frac{1}{2}-\frac{1}{2}\nu}\delta _{\nu 0}\, ,
\end{equation}
 where \( N_{jj^{\prime }n} \) is a normalizing factor which turns out to be
\( \frac{\sqrt{2j+1}\sqrt{2j^{\prime }+1}}{4\pi }(-1)^{j-\frac{1}{2}+n} \)
{[}\ref{we2}{]}. In other words, only the factor \( c_{n0}(jj^{\prime }) \)
enters into our expression for the efficiency operator.

A formal derivation of this result is given in our paper under reference {[}\ref{we2}{]}.
However, that derivation does not quite relate to the attenuation properties
of a detector. On the other hand, we feel that our present approach is physically
more transparent. It also throws some light on a couple of questions of fundamental
epistemological interest. Does the detection process have a role in this type
of experiment? If so, what is that role like? Obviously, such questions can
be important from the standpoint of the theory of measurement.

Calculation of the angular correlation function can now go through as in reference
{[}\ref{we2}{]}. We finally get

\begin{equation}
\label{11}
\begin{array}{ccc}
W(\theta ) & = & \sum _{k}z_{k}(1)z_{k}(2)(-1)^{j_{1}+j_{2}}c_{k0}(j_{1}j^{\prime }_{1})c^{\star }_{k0}(j_{2}j^{\prime }_{2})\\
 &  & \times w(J_{b}J^{\prime }_{b}j_{1}j^{\prime }_{1},kJ_{a})w(J_{b}J^{\prime }_{b}j_{2}j^{\prime }_{2},kJ_{c})P_{k}(cos\theta )\, ,
\end{array}
\end{equation}
 where the \( w \)'s are Racah coefficients, \( j_{1} \) and \( j_{2} \)
are the angular momenta of the photo- and Auger electrons, \( J_{a},J_{b},J_{c} \)
the angular momenta of the atom in its initial, intermediate and final states
respectively. Here \( k \) is an even integer ranging from \( 0 \) to \( k_{max} \),
\( k_{max} \) being defined as follows. Let \( \left\{ \left\{ j_{1}+j^{\prime }_{1}\right\} _{max},\left\{ j_{2}+j^{\prime }_{2}\right\} _{max}\right\} _{min}=p \).
Then

\begin{equation}
\label{12}
\begin{array}{ccc}
k_{max} & = & p\, \, if\, p\, is\, even,\\
 & = & (p-1)\, if\, p\, is\, odd.
\end{array}
\end{equation}
 The set of primed angular momentum quantum numbers represent virtual states
which may arise from possible interaction with other atoms and electrons.

To sum up, our main finding in this paper is that the angular correlation function
\( W(\theta ) \) arises directly from an unavoidable indeterminacy in the actual
directions of emission of the two electrons. From our analysis above, it should
be clear that this is not a matter of technical imperfection but a basic
restriction imposed on the detection process by the dynamical nature of the
problem.

Our results for the double photoionization of xenon are discussed in reference
{[}\ref{we2}{]} in some detail.

One of the authors (CS) is indebted to the University Grants Commission of India
for support in the form of a junior research fellowship.

\vspace{0.3cm}
{\par\centering \resizebox*{!}{0.5\textheight}{\rotatebox{-90}{\includegraphics{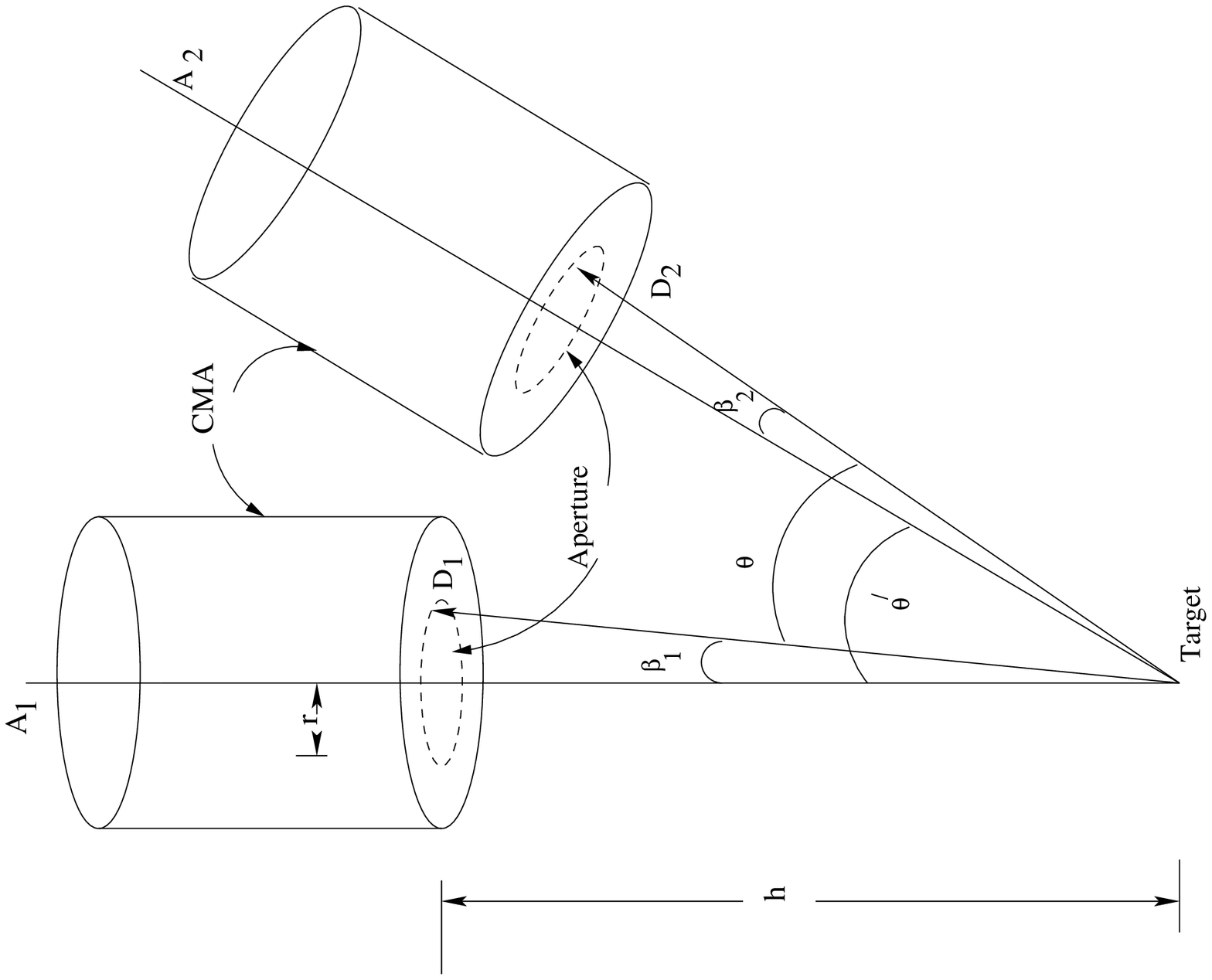}}} \par}
\vspace{0.3cm}

{\par\centering Fig 1 : Geometrical arrangement of the detectors\par}

\end{document}